\documentclass[aps, prb, superscriptaddress, twocolumn, letterpaper, 10pt, amsfonts, amsmath, amssymb, groupedaddress, longbibliography,  floatfix]{revtex4-2}\usepackage{graphicx} 

\usepackage{physics}
\usepackage{float}
\usepackage{xcolor}
\usepackage{soul}
\usepackage[normalem]{ulem}
\usepackage[mathscr]{euscript}
\usepackage{bm}
\usepackage{dsfont}
 
\usepackage[colorlinks=true,citecolor=blue,urlcolor=blue,linkcolor=blue]{hyperref}
\usepackage{float}

\newcommand{\ii}{\mathrm{i}}
\newcommand{\ee}{\mathrm{e}}
\renewcommand{\Vec}[1]{\bm{#1}}
\newcommand{\hc}{{\rm{H.c.}}}

\newcommand{\B}{{\rm{B}}}

\newcommand{\figref}{Fig.~\ref}

\renewcommand{\eqref}[1]{{Eq.~(\ref{#1})}}

\begin{document}
\title{
Microscopic solutions for vortex clustering in two-band  type-1.5 superconductors}

\author{Igor Timoshuk} 
\affiliation{Department of Physics, The Royal Institute of Technology, Stockholm SE-10691, Sweden}
\affiliation{Wallenberg Initiative Materials Science for Sustainability, Department of Physics, The Royal Institute of Technology, Stockholm SE-10691, Sweden}
\author{Egor Babaev} 
\affiliation{Department of Physics, The Royal Institute of Technology, Stockholm SE-10691, Sweden}
\affiliation{Wallenberg Initiative Materials Science for Sustainability, Department of Physics, The Royal Institute of Technology, Stockholm SE-10691, Sweden}

\begin{abstract}
Two-band superconductors exhibit a distinct phase characterized by two correlation lengths, one smaller and the other larger than the magnetic field penetration length. This regime was coined type-1.5 superconductivity, with several unconventional properties, such as vortex clustering. However, a fully microscopic solution for vortex clusters has remained challenging due to computational complexities beyond quasiclassical models. This work presents numerical solutions obtained in a fully self-consistent two-band Bogoliubov–de Gennes model.We show the presence of discrepant correlation lengths leading to vortex clustering in two-band superconductors.
\end{abstract}

\maketitle

The original work by Ginzburg and Landau introduced the concept of coherence length $\xi^{GL}$ \cite{landau1950k} and classified superconductors by a single Ginzburg-Landau parameter: the ratio $\kappa^{GL}=\lambda/\xi^{GL}$ of two fundamental length scales---the magnetic field penetration length $\lambda$ and the coherence length $\xi^{GL}$. The latter is a fundamental length scale that governs the asymptotic behavior of the modulus of superconducting gap $|\Delta|e^{i\theta}$ or equivalently, up to a different prefactor \cite{gor1959microscopic}---the Ginzburg-Landau order parameter field $|\Psi|e^{i\theta}$.
The existence of this order parameter and, hence, the fundamental length $\xi^{GL}$ is guaranteed by the fact that a superconductor breaks local $U(1)$ symmetry \cite{landau1950k}. 
Within Ginzburg-Landau's (GL) theory, the superconductor allows
repulsively interacting vortices for $\kappa^{GL}>1/\sqrt{2}$, called type-2 superconductivity \cite{abrikosov1957magnetic}. The vortices form lattices 
when the magnetic field is larger than the first critical magnetic field $H_{c1}$
and smaller than the second critical magnetic field $H_{c2}$ \cite{abrikosov1957magnetic}. 
These critical magnetic fields were introduced in the series of experimental works by Shubnikov \textit{et al.} \cite{rjabinin1935magnetic,shubnikov1937magnetic}, so the state
forming $H_{c1}<H<H_{c2}$ is also referred to as Shubnikov's phase.
In what follows, we will absorb the factor $\sqrt{2}$ into the definition of coherence length $\xi\equiv \sqrt{2}\xi^{GL}$. Thus the GL criterion for type-2 superconductivity in these notations is $\kappa\equiv \sqrt{2}\kappa^{GL}>1$.
Each vortex has a current-carrying area of radius $\lambda$ around its core. At the vortex core, the modulus of the order parameter is suppressed.  
The exact definition of $\xi$ is the characteristic exponent of the gap decay far away from the center of the vortex core. At low temperatures, the overall size of the vortex core is smaller than $\xi$ in the simplest models \cite{gygi1990ang}.
Note also that, in strongly type-2 superconductors, the long-range asymptotic of a vortex is affected by nonlinearities \cite{plohr1981behavior}.
When $\kappa<1$, in an ordinary Ginzburg-Landau theory, the vortex is too energetically expensive and thermodynamically unstable. At $\kappa=1$ vortices do not interact \cite{kramer,saint1969type,bogomol1976stability} \footnote{Going beyond Ginzburg-Landau theory, at low temperatures, one obtains microscopic corrections that 
result in small nonmonotonic intervortex forces of very different origin and form
compared to the current paper (i.e. not being driven by overlaps
of outer cores). According to the microscopic theory \cite{jacobs,leung1,eilenberger,weber1987phase,klein,miranovic2003thermodynamics} the effect occurs in a narrow range of $\kappa^{GL} \leq 1.1$.
It should be noted that it is a small microscopic model-specific effect beyond Ginzburg-Landau theory, i.e., it cannot be described by carrying Ginzburg-Landau expansion to higher order in $\tau=(1-T/T_c)$  \cite{rybakov2021absence}.}.

Ginzburg and Landau's work \cite{landau1950k} classified single-component superconductors. Today, many superconducting states of interest break multiple symmetries, for example, featuring a breakdown of time-reversal symmetry translation symmetry or nematicity. Therefore, they require description in terms of multiple order parameter fields $|\Psi_i|e^{i\theta_i}$ and must be characterized by multiple coherence lengths.
It was pointed out in \cite{Babaev.Speight:05} that, in multicomponent systems, a new regime is possible where some coherence lengths are shorter than the magnetic field penetration length and some are larger: $\xi_1<\xi_2<...\lambda <\xi_n<\xi_{n+1}...$.
This regime was termed type-1.5 in \cite{moshchalkov}.
The vortex excitations there can be viewed as composite objects. Namely, they are bound states of elementary vortices
with phase winding only in one of the components $\oint d l\nabla\theta_n=\pm 2\pi$.
Such elementary constituents carry a fraction of flux quantum and are much more energetically expensive than integer-flux vortices \cite{babaev2002vortices}.
Then, in an external field, the system is expected to form composite vortices where all components have phase winding around the common core, which can also be viewed as bound states of fractional vortices where fractions add up to one flux quantum.
The fractional vortices
with phase winding only in the single band have been recently experimentally observed \cite{iguchi2023superconducting}.
When some coherence lengths are larger than the magnetic field penetration length, the density-density interaction results in long-range attractive intervortex forces. Meanwhile, the magnetic- and current- interaction gives short-range repulsion \cite{Babaev.Speight:05,johan1,johan2}. Consequently, in a low magnetic field, such a system exhibits vortex clustering and phase separation in vortex droplets and the Meissner domains \cite{Babaev.Speight:05,johan1,johan2,nonpairwise,diaz2017glass}.
The concept of type-1.5 superconductivity was also generalized to other systems beyond superconductivity, such as the typology of
quantum Hall systems \cite{parameswaran2012typology} and neutron stars \cite{wood2022superconducting}.

In the above, we emphasized the case of multiple broken symmetries.
The more nontrivial case for typology is represented by the commonly occurring superconducting materials with multiband electronic structures. Such systems have multiple superconducting gaps forming on different bands $|\Delta_\alpha|e^{i\theta_\alpha}$
but where interband Josephson interaction breaks symmetry down to single $U(1)$. In that case, symmetry does not guarantee multiple correlation lengths, even with multiple bands. Nonetheless, in the simplest Landau theories, explicit symmetry breaking does not prohibit the existence of extra coherence length.
It was discussed at the level of two-band Ginzburg-Landau theory
in \cite{johan1,johan2,nonpairwise,Johan} that in multiband $U(1)$ systems, multiple coherence lengths arise and are associated with different linear combinations of the gaps fields \footnote{Note that in general, the inclusion of anisotropies creates multiple scales in the magnetic field penetration lengths too \cite{winyard2019hierarchies}}. However, the justification of multiple coherence lengths is nontrivial: the two-band Ginzburg-Landau model is an expansion in multiple small parameters associated with multiple small gaps and corresponding gradients. Such an expansion is not always justified as it is based not on a small parameter guaranteed by symmetry but depends on the structure of the intercomponent interaction \cite{silaev2}, such as strength and presence or absence of frustration. The conditions and parameter range where two coherence lengths occur in two-band $U(1)$ systems were studied in microscopic quasiclassical Eilenberger formalism in \cite{silaev1,silaev2}, confirming, at the level of quasiclassical theory the existence of length scale hierarchy $\xi_1 <\lambda <\xi_2$ in two-band systems that break only a single symmetry. The simplest two-band models require weak interband coupling to realize this regime \cite{silaev1,silaev2}. These works also calculated asymptotic intervortex forces in two-band Eilenberger formalism. 
Solutions for vortex clusters in the type-1.5 regime were obtained in several microscopically derived Ginzburg-Landau models \cite{silaev2,Garaud.Corticelli.ea-2018a,agterberg2014microscopic}.
However, to date, no solutions for vortex clusters in the type-1.5 regime were obtained in microscopic models. 
Vortex clusters were observed experimentally in several multiband systems and attributed to type-1.5 physics in 
\cite{moshchalkov,moshchalkov2,Ray.Gibbs.ea:14,biswas2020coexistence,bolotina2022checkerboard,ge2022experimental,bolotina2023low,curran2023search}.

A microscopic approach that retains even the shortest-length-scales physics is the
Bogoliubov-de Gennes (BdG) formalism \cite{de2018superconductivity}.
Within this formalism, fully microscopic solutions, including self-consistent calculation of magnetic field, were obtained for an isolated Abrikosov vortex in \cite{gygi1990tun, gygi1990ang, gygi1991elstr}. However, obtaining the solutions for vortex clusters is significantly more challenging as one cannot rely on an axially symmetric ansatz.
Here, we report solutions of vortex clusters in the fully self-consistent numerical treatment of the multiband Bogoliubov--de Gennes model, including a self-consistent solution for the magnetic field.

The two-band BdG model that we consider is defined on a two-dimensional square lattice, described by the mean-field Hamiltonian 
\begin{equation}\label{eq: mfHamiltonian}
\begin{gathered}
    \mathcal{H}= -  \sum_{\sigma\alpha}\sum_{<ij>} e^{\ii q A_{ij} } c^\dagger_{i \sigma \alpha } c_{j\sigma \alpha} +\\
 + \sum_{i \alpha} \left( \Delta_{i \alpha} c^\dagger_{\uparrow i \alpha} c^\dagger_{\downarrow i \alpha} +  \hc \right) + \frac{1}{2}F_{m}\,. 
\end{gathered}
\end{equation}
Here $<ij>$ denotes all nearest neighbor pairs, $c_{i \sigma \alpha}$ is the fermionic annihilation operator at position $i$, with spin $\sigma$ ($\sigma \in \lbrace \uparrow, \downarrow \rbrace$) and band index $\alpha$ ($\alpha \in \lbrace 1,2 \rbrace$) and $\hc$ denotes Hermitian conjugation. The phase factor $\exp(\ii q A_{ij})$ accounts for the interaction with the magnetic vector potential $\Vec{A}$ through Peierls substitution \cite{Peierl,feynman2011mainly},
$\frac{1}{2}F_{m}$ is the magnetic field energy density.
\begin{equation}
    A_{ij} = \int_j^i \Vec{A} \cdot \dd{\Vec{\ell}}\,,
\end{equation}
the gap fields are defined as
\begin{equation}\label{eq: selfConsistency}
    \Delta_{i \alpha} = \sum_\beta V_{\alpha \beta} \expval{c_{\uparrow i \beta} c_{\downarrow i \beta}}\,,
\end{equation}
where $V_{\alpha \beta} = V_{\beta \alpha}^*$ stands for quadratic interaction term, and the fermionic current is
\begin{equation}\label{eq:curr}
    J_{ij} = -2q\sum_{\alpha \sigma} \mathrm{Im}\left(\langle c_{i\alpha\sigma}^\dagger c_{j\alpha\sigma}\rangle e^{\ii q A_{ij}} \right) 
\end{equation}
and discrete version of Maxwell's equation $\curl \curl \Vec{A} = \Vec{J}$   determines the connection between $A_{ij}$ and $ J_{ij}$. 

The free energy associated with the tight-binding Hamiltonian \eqref{eq: mfHamiltonian} may be expressed as 
\begin{equation}\label{eq:Fsum}
\begin{gathered}
    F_H = \sum_{i}\Vec{\Delta}_i^\dagger V^{-1}\Vec{\Delta}_i -k_\B T\Tr\ln\qty(\ee^{-\beta \mathcal{H}}+1) + \\
    + \frac{1}{2}\sum_{\mathrm{plaquettes}} B^2,
\end{gathered}
\end{equation}
where the magnetic field $B=\curl \Vec{A}$ is defined on plaquettes. 

The self-consistency equations \eqref{eq: selfConsistency}, \eqref{eq:curr}, along with the Maxwell equation, are solved numerically using an iterative scheme, described in \cite{benfenati2023}. 
Two independent codes were used to validate the solutions.
New values are obtained for the vector potential and the gaps during each iteration, using \eqref{eq: selfConsistency}. They can be calculated by obtaining the eigenvectors $c_{\sigma i \alpha}$ by directly diagonalizing the Hamiltonian or using the Chebyshev spectral expansion scheme.
The key results of the paper are obtained by
graphic processing units (GPU)-based exact matrix diagonalization, used for free energy calculation \eqref{eq:Fsum}. 
In addition, we report approximate solutions for a larger vortex cluster obtained using the approximate Chebyshev spectral expansion method for larger vortex clusters.
The iteration procedure stops when the convergence criteria $\abs{\delta p/\left(p +\epsilon\right)} < \varepsilon$ is achieved for each of the parameters $\Delta_1$, $\Delta_2$, and $A$ simultaneously. Note that we do not calculate stray fields outside the sample, the model may be interpreted as a part of a stack of two-dimensional lattices.

Below, we report microscopic solutions for
vortex clusters in the BdG model. We demonstrate (i) the existence of multiple correlation lengths in the microscopic solution, (ii) that these length scales form the required hierarchy: $\xi_1<\lambda <\xi_2$, (iii) the intervortex interaction potential has a minimum at a finite distance, and (iv) a multi-quanta vortex separates into a bound state of single-quanta vortices, forming a cluster. 

We study a square sample with linear size $L$ and open boundary conditions.
The crucial aspect is to avoid mesoscopic effects on vortex physics. For that reason, the vortices are initiated with an initial guess far away from sample boundaries, and the sample is chosen to be significantly large to avoid vortex escape due to boundary attraction.
For the same reason, only the regime with moderate disparity of the length scales could be studied to have all the characteristic length scales much smaller than the grid size and, simultaneously, significantly larger than lattice spacing.
Hence, our choice of parameters is motivated by computational constraints rather than the physics of the concrete compound.
The calculations based on the Chebyshev approximation method were performed on a grid with $L=64$. We used the exact diagonalization method with double precision for free energy calculations, so the system size is decreased to $L=48$ sites.
Since the quasiclassical analysis \cite{silaev1} suggests that interband coupling should be very weak to have well-defined multiple correlation lengths, we analyze the Hamiltonian \eqref{eq: mfHamiltonian} with the following numerical parameters $q=0.6$, $V_{11}=2.8$, $V_{22}=2.2$, $V_{12}=0.01$, and $T=0.264$ in units of the bandwidth.
We use the convergence criteria $\epsilon=10^{-8}$ and $\varepsilon=10^{-6}$.

\begin{figure}[t]
    \centering
    
    \includegraphics[width=\columnwidth]{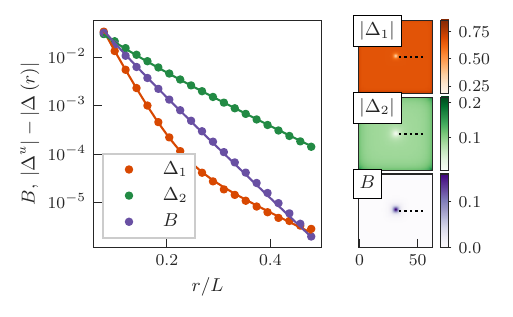}
    \caption{Absolute values for superconducting gap and magnetic field for single vortex, showing three distinct length scales. Left: asymptotic of single-vortex solution. Dots represent numerical data for the gap amplitudes $\Delta_{1,2}$ and magnetic field $B$; lines show the fit for the function \eqref{eq:jos_gaps}. Right: heat maps for $\left| \Delta_1 \right|$,  $\left| \Delta_2 \right|$ and  $\left| B \right|$. Dot lines show the cross section line, represented on the left figure.} \label{fig: xi_12_1}
\end{figure}

First, we analyze the structure of a single vortex. 
Single-vortex states were calculated using the Chebyshev spectral expansion approximation. From these solutions, the asymptotics for magnetic field and order parameter correlation length were obtained \figref{fig: xi_12_1}.  

In the presence of even a tiny inter-band Josephson coupling $V_{12}$, coherence lengths are affected quantitatively and qualitatively. 
Calculations in two-band Ginzburg-Landau  \cite{johan1,johan2} and Eilenberger  \cite{silaev1,silaev2} formalisms
predict that, away from a vortex, each
gap field approaches its asymptotic value $\left|\Delta_{1,2}^\mathrm{u}\right|$ with two length scales. In these models, the gap asymptotic is given by a combination of two modified Bessel functions: 
\begin{equation}
\begin{aligned} \label{eq:jos_gaps}
    \left|\Delta_1 \left(r\right) \right| = \left|\Delta^\mathrm{u}_1\right| - q_1 \cos\Theta K_0 \left(r/\xi_1\right) + q_2 \sin\Theta K_0 \left(r/\xi_2\right) \\
    \left|\Delta_2 \left(r\right) \right| = \left|\Delta^\mathrm{u}_2 \right|- q_1 \sin\Theta K_0 \left(r/\xi_1\right) - q_2 \cos\Theta K_0 \left(r/\xi_2\right)
\end{aligned}
\end{equation}
This implies that due to interband coupling, the coherence lengths $\xi_{1,2}$ are associated with the linear combination of the gap fields rather than individual bands.

The cross section on \figref{fig: xi_12_1} shows two length scales in the gap function. 
The solution shows that even a weak interband coupling forces the same long-range asymptotic on both gap functions despite very different behavior of $\Delta_1$ near the origin.
The solutions can be fitted with \eqref{eq:jos_gaps}, which gives us $\xi_1\approx 1.27$, $\xi_2 \approx 4.53$, $q_1 \approx 0.043$, $q_2 \approx  -0.48$, and $\Theta \approx 0.495\pi$; meanwhile, the fit for $B=q_be^{-x/\lambda}\sqrt{\lambda/x}$ gives $\lambda \approx 2.25$ and $q_b=0.14$. Therefore, it shows that $\xi_1<\lambda<\xi_2$, verifying that the system is in the type-1.5 regime for these coupling constants.  

Upon establishing the type-1.5 hierarchy of the length scales, we demonstrate that the system forms vortex clusters due to the competition of long-range attractive core-core interaction set by coherence length $\xi_2$ and shorter-range current-current and magnetic interaction set by the magnetic field penetration length $\lambda$.
In the next step, we analyze the interaction energy between two vortices.
To ensure that the existence of nonmonotonic intervortex forces are not artifacts of numerical grid pinning the calculations are performed for a variety of initial conditions. The total free energy of the system  \eqref{eq:Fsum} is orders of magnitude higher than the intervortex interaction energy, so the exact diagonalization method and double precision were used to ensure the accuracy of the energy calculations. 

\begin{figure}[t]
    \centering
    
    \includegraphics[width=\columnwidth]{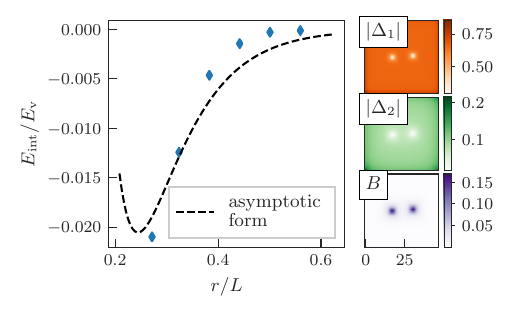}
    \caption{Left: vortex pair interaction energy \eqref{eq:F_calc}, expressed in single-vortex energy units $E_{\mathrm{v}}=F_{\mathrm{v}} - F_{\mathrm{u}}$ vs relative distance $r/L$ between the vortex cores.
    The convergence at blue points on the left panel is achieved due to numerical grid pinning. The leftmost data point corresponds to the calculated minimum of the interaction energy. The dashed line is the analytic fit based on \eqref{eq:E_int}.
    Right:  $\left| \Delta_1 \right|$,  $\left| \Delta_2 \right|$ and  $\left| B \right|$ for the minimal energy solution. There is a noticeable core overlap in $\Delta_2$.} \label{fig: en_dep}
\end{figure}

If the distance between the vortices in the initial guess is large enough, the interaction force will not move the vortex from its initial position
due to the exponential falloff of interaction,
and the presence of an underlying lattice pinning.
This allows us to calculate the energy of vortex interaction directly on different distances away from the minimum of the potential \figref{fig: en_dep}. The interaction energy of vortex pair $E_{\mathrm{int}}\left(r\right)$ is calculated as follows.
We calculate the energy of a vortex pair with the positions $\textbf{r}_1$ and $\textbf{r}_2$: $F_\mathrm{p}\left(L, \textbf{r}_1, \textbf{r}_2\right)$; from this we subtract (i) the energy of the system in the absence of vortices $F_\mathrm{u}\left(L\right)$ and
(ii) the energies of the solutions for single vortices, which are calculated at the same positions as vortices in the pair  $\textbf{r}_{1,2}$, in order to diminish the finite-size and discretization effects and to offset the energies of the vortex pinning by numerical grid. However, the grid pinning means that we slightly overestimate the interaction energy, compared to the analytic expression (\ref{eq:E_int}),  due to nonlinear corrections,
\begin{equation} \label{eq:F_calc}
    E_{\mathrm{int}}\left(r\right) = F_\mathrm{p}\left(L, \textbf{r}_1, \textbf{r}_2\right) + F_{\mathrm{u}}\left(L\right) - F_\mathrm{v}\left(L, \textbf{r}_1\right) - F_\mathrm{v}\left(L, \textbf{r}_2\right) ,
\end{equation}
where $r=\abs{\textbf{r}_1-\textbf{r}_2}$. 

The long-range asymptotic form for the vortex-vortex interaction energy a in type-1.5 superconductor calculated in
continuum two-band Ginzburg-Landau  \cite{johan1,silaev2} and Eilenberger  \cite{silaev1} formalisms 
  has the form 
\begin{equation} \label{eq:E_int}
    V\left(r\right) = \alpha\left( q_b^2 K_0\left(r / \lambda\right) - q_2^2 K_0\left(r / \xi_2\right) - q_3^2 K_0\left(r / \xi_1\right) \right),
\end{equation}
where $\alpha$ is a positive constant \footnote{In more general models, the prefactors can have additional parameters \cite{johan2} and more terms can be present \cite{Johan,winyard2019hierarchies}} This is different from the monotonic interaction potential in a standard single-component Ginzburg-Landau model \cite{kramer}.

The two dominant terms with $\lambda$ and $\xi_2$ in this expression give a minimum interaction potential at a certain intervortex distance that depends on the competing coherence and magnetic field penetration length scales. 
We can compare this approximate expression with the results of our BdG-based calculations. First, since we are interested only in long-range forces, we omit the part with the shortest length scale $~ K_0\left(r / \xi_1\right)$
from \eqref{eq:jos_gaps}. We extracted the coherence and magnetic field penetration lengths from our solutions in the above. Using these lengths, we fit the results from \eqref{eq:F_calc} using \eqref{eq:E_int} and obtain $\alpha$.
Although strictly speaking, the form \eqref{eq:E_int} is derived using different continuum models, that equation approximately fits the calculated intervortex potential.
It also shows that the long-range attraction is dominated by density-density interaction between the extended vortex cores, and short-range repulsion is dominated by current-current and magnetic interactions.

\begin{figure}[t]
    \centering
    \includegraphics[width=1.0\columnwidth]{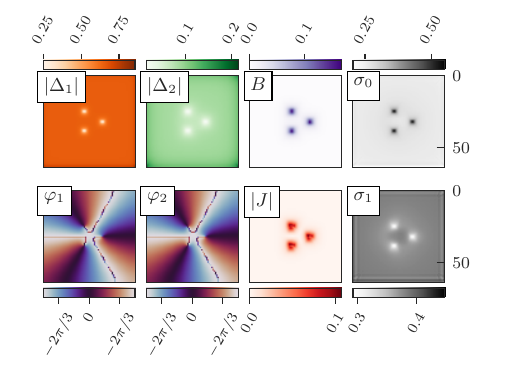}
    \caption{Cluster made of three vortices. The panels show the absolute values $\abs{\Delta_{1,2}}$ and phases $\varphi_{1,2}$ for gap distributions, magnetic field $B$, and current $\abs{J}$ from \eqref{eq:curr} in the system and tunneling conductance below $\sigma_0$ and above $\sigma_1$ the gap. The initial state of the system was a single vortex with charge 3. There is a visible overlap of the vortex cores in the second component. The distance between the cores for each pair of vortices is the same as for the minimal energy solution on \figref{fig: en_dep}.} \label{fig: cl_3}
\end{figure}

Finally, an approximate solution for a larger vortex cluster is obtained. 
The exact diagonalization methods are computationally expensive and larger system sizes are necessary to study multivortex clusters, so the Chebyshev spectral expansion method was used.
The system \eqref{eq: mfHamiltonian} is evaluated for a single multiquanta vortex solution as an initial condition. 
The giant vortex is not a stable configuration, and it decays into a vortex cluster. On \figref{fig: cl_3} is shown the solution for an initial guess of the giant vortex with charge 3.

In conclusion, in a fully microscopic formalism, we demonstrated superconductivity beyond the type-1/type-2 dichotomy in two-band superconductors. Such superconductors break only a single symmetry due to interband coupling.
Nonetheless, when the interband coupling is weak, the obtained vortex solutions clearly show the effects of multiple correlation lengths.
The numerical solutions show that these correlation lengths are hybridized, i.e., associated with different linear combinations of the gap fields. 
For weak interband coupling, we find 
hierarchy of the length scales
$\xi_1<\lambda<\xi_2$ that leads
to attractive intervortex interaction at large separation due to core-core overlap.
It does not exclude large disparity of coherence lengths at stronger interband coupling, for example, in the cases of frustrated interband coupling or proximity to a phase transition into a superconducting state with different symmetry \cite{Johan,Garaud.Corticelli.ea-2018a}.
The Bogoliubov--de Gennes formalism also allows us to calculate signatures of the vortex clusters in scanning tunneling microscopy. In such a probe, a vortex cluster in the considered microscopic model can appear as a group of vortices with individual small cores yet having a significant attractive interaction. 
The computational complexity of BdG models limits system sizes and, therefore, coupling constants that we can consider now. An interesting further direction could be solutions and STM signatures for material-specific small coupling constants and larger numerical grids.

\begin{acknowledgments}
We thank A. Talkachov for insightful discussions and for sharing his exact diagonalization CUDA code. This work was supported by 
the Swedish Research Council Grants No. 2022-04763, by Olle Engkvists Stiftelse, and the Wallenberg Initiative Materials Science for Sustainability (WISE) funded by the Knut and Alice Wallenberg Foundation.
\end{acknowledgments}

\bibliography{bibliography.bib}
\end{document}